\documentclass[notitlepage]{revtex4-1}

\usepackage[utf8]{inputenc}
\usepackage[colorlinks=true,citecolor=blue,linkcolor=blue]{hyperref}
\usepackage[normalem]{ulem}
\usepackage{url}
\usepackage{graphicx,wrapfig,float,slashed,cancel}
\usepackage{amsmath,amssymb,epsfig,graphicx,xcolor,stmaryrd}
\usepackage{bm}
\usepackage{enumitem}
\usepackage{multirow}

\definecolor{darkblue}{RGB}{1, 90, 173}

\begin{document}


\title{Investigation of a candidate spin-$\frac{1}{2}$ hidden-charm triple strange pentaquark state $P_{csss}$}

\author{K.~Azizi}
\email{ kazem.azizi@ut.ac.ir}
\thanks{Corresponding author}
\affiliation{Department of Physics, University of Tehran, North Karegar Avenue, Tehran
14395-547, Iran}
\affiliation{Department of Physics, Do\v{g}u\c{s} University, Dudullu-\"{U}mraniye, 34775
Istanbul, Turkey}
\affiliation{School of Particles and Accelerators, Institute for Research in Fundamental Sciences (IPM) P.O. Box 19395-5531, Tehran, Iran}
\author{Y.~Sarac}
\email{yasemin.sarac@atilim.edu.tr}
\affiliation{Electrical and Electronics Engineering Department,
Atilim University, 06836 Ankara, Turkey}
\author{H.~Sundu}
\email{ hayriye.sundu@kocaeli.edu.tr}
\affiliation{Department of Physics, Kocaeli University, 41380 Izmit, Turkey}

\date{\today}

\preprint{}

\begin{abstract}

A candidate triple strange pentaquark state, $P_{csss}$, is investigated through its strong decay channel $P_{csss} \rightarrow \Omega^-J/\psi $. To calculate the relevant strong coupling constants, two possible interpolating currents with spin-parity $J^P=\frac{1}{2}^{-}$ are used. Though the chosen currents for the  state under consideration have spin-parity quantum numbers $J^P=\frac{1}{2}^{-}$,  they  couple to both the positive and negative parity states simultaneously and the corresponding decay widths are obtained for both parities. These widths are obtained as   $\Gamma(P_{csss} \rightarrow J/\psi \Omega^-)=201.4\pm 82.5~\mathrm{MeV}$  for the negative and  $\Gamma(\widetilde{P}_{csss} \rightarrow J/\psi \Omega^-)=316.4\pm 107.8~\mathrm{MeV}$ for the positive parity state when the first current is used.  For the second current, we obtain $\Gamma(P_{csss} \rightarrow J/\psi \Omega^-)=252.5\pm 116.7~\mathrm{MeV}$ for the negative and  $\Gamma(\widetilde{P}_{csss} \rightarrow J/\psi \Omega^-)=361.1\pm 98.4~\mathrm{MeV}$ for the positive parity state. These results may provide insights into future experimental observations of such candidate states and help to distinguish and fix their properties.

\end{abstract}


\maketitle

\renewcommand{\thefootnote}{\#\arabic{footnote}}
\setcounter{footnote}{0}
\section{\label{sec:level1}Introduction}\label{intro} 

The past few decades have become the era of the observation of various hadrons which include both the conventional hadrons with either three-quark or a quark-antiquark substructures and their excitations as well as non-conventional exotic states. As a result, nowadays, investigation of the properties of the exotic states, such as their substructures and quantum numbers, has become an attractive research field. Since their first proposal by Gell-Mann~\cite{Gell-Mann}, their existence and possible properties were widely probed out theoretically until their first observation was announced for $X(3872)$ in 2003 by the Belle Collaboration~\cite{Choi:2003ue}. This observed state was a tetraquark state, and later other collaborations such as the CDF II Collaboration~\cite{CDF:2003cab,CDF:2009nxk}, the BABAR Collaboration~\cite{BaBar:2004oro}, D0 Collaboration~\cite{D0:2004zmu}, LHCb Collaboration~\cite{LHCb:2011zzp}, and CMS Collaboration~\cite{CMS:2013fpt} validated this observation.     

Another member of the exotic state family is the pentaquark configuration, for which the first observation came to light in 2015 in the $J/\psi+p$ decay channel by the observation of the LHCb Collaboration~\cite{Aaij:2015tga}. The observed states were stated to be $P_c(4380)$ and $P_c(4450)$. However, in 2019 the updated analyses of the LHCb Collaboration indicated the splitting of the $P_c(4450)$ into two states, labeled as $P_c(4440)$ and $P_c(4454)$~\cite{Aaij:2019vzc}. In Ref.~\cite{Aaij:2019vzc} a new state, $P_c(4312)$, was also announced. The resonance parameters for these states were reported as \cite{Aaij:2015tga,Aaij:2019vzc}:
\begin{eqnarray}
&& m_{P_c(4380)^+}=4380 \pm 8 \pm 29~\mbox{MeV} ~~~~~~~~~~ \Gamma_{P_c(4380)^+}=205 \pm 18 \pm 86~\mbox{MeV}, \nonumber\\
&& m_{P_c(4440)^+}=4440.3 \pm 1.3 ^{+ 4.1}_{-4.7}~\mbox{MeV} ~~~~~~~ \Gamma_{P_c(4440)^+}= 20.6 \pm 4.9^{+8.7}_{-10.1}~\mbox{MeV},\nonumber\\
&& m_{P_c(4457)^+}=4457.3 \pm 0.6 ^{+ 4.1}_{-1.7}~\mbox{MeV} ~~~~~~~ \Gamma_{P_c(4457)^+}= 6.4 \pm 2.0^{+5.7}_{-1.9}~\mbox{MeV},\nonumber\\
&& m_{P_c(4312)^+}=4311.9 \pm 0.7^{ +6.8}_{-0.6}~\mbox{MeV} ~~~~~~~ \Gamma_{P_c(4312)^+}=9.8 \pm 2.7 ^{ +3.7}_{-4.5}~\mbox{MeV}.\nonumber
\end{eqnarray}
These observations were followed by the report of two more observations in 2021 and 2022 which are $P_{cs}(4459)$~\cite{LHCb:2020jpq} and $P_c(4337)$~\cite{LHCb:2021chn}, respectively. The $P_{cs}(4459)$ contains a strange quark and was observed in $J/\psi\Lambda$ invariant mass spectrum, whereas $P_c(4337)$ was observed in both $J/\psi p$ and $J/\psi \bar{p}$ invariant mass spectra with the following spectral properties \cite{LHCb:2020jpq,LHCb:2021chn}:
\begin{eqnarray}
&& m_{P_{cs}(4459)^0}=4458.8 \pm 2.9^{+4.7}_{-1.1}~\mathrm{MeV}~~~~~~ \Gamma_{P_c(4459)^0}=17.3 \pm 6.5^{+8.0}_{-5.7}~\mathrm{MeV}, \nonumber\\
&& m_{P_{cs}(4337)^+}=4337^{+7}_{-4}{}^{+2}_{-2}~\mathrm{MeV}~~~~~~~~~~~~~~ \Gamma_{P_c(4337)^+}= 29 ^{+26}_{-12}{}^{+14}_{-14}~\mathrm{MeV}.\nonumber
\end{eqnarray}
By the ripple of excitement brought by these observations, these hadrons have become one of the focus of intense theoretical studies. These theoretical studies aimed to investigate the properties of these states to understand their natures and substructures. Besides, some of these works focused on other possible new states to contribute to experiments by offering new possible states for their future observations. Via various approaches and different structure assumptions, the observed states were investigated thoroughly. In some of these works, to shed light on their obscure substructure, these pentaquarks were treated as compact pentaquark states. Diquark-diquark-antiquark or diquark-triquark forms were adopted in Refs.~\cite{Wang:2015epa,Maiani:2015vwa,Anisovich:2015zqa,Li:2015gta,Lebed:2015tna,Anisovich:2015cia,Wang:2015ava,Wang:2015ixb,Ghosh:2015ksa,Wang:2015wsa,Wang:2016dzu,Zhang:2017mmw,Giannuzzi:2019esi,Wang:2019got,Wang:2020rdh,Ali:2020vee,Azizi:2021utt,Wang:2020eep,Azizi:2021pbh,Zhu:2015bba,Gao:2021hmv}. Refs.~\cite{Guo1,Mikhasenko:2015vca,Liu1,Bayar:2016ftu,Guo2} considered the possibility of their arising from the kinematical effects. Another common structure suggested for the pentaquark states is the molecular form. Based the closeness of their observed masses to meson baryon thresholds and their small widths, they were investigated using different theoretical methods adopting molecular structures. Among these theoretical models considering molecular interpretations are the one-boson exchange potential model ~\cite{Wang:2021hql,Chen:2022onm,Yalikun:2021dpk,Yalikun:2021bfm,Chen:2021kad,Wang:2021hql,Pan:2020xek,Liu:2019zvb,Wang:2019nwt,Chen:2019asm,Wang:2019aoc,Chen:2016ryt,Chen:2016heh}, quasipotential Bethe-Salpeter~\cite{He:2019ify,He:2019rva,Zhu:2021lhd}, the contact-range effective field theory~\cite{Liu:2019tjn,Liu:2020hcv,Peng:2020gwk,PavonValderrama:2019nbk}, the effective Lagrangian approach~\cite{Cheng:2021gca,Ling:2021lmq,Xiao:2019mvs,Lu:2016nnt}, and the QCD sum rule method~\cite{Chen:2016otp,Azizi:2016dhy,Azizi:2018bdv,Azizi:2020ogm,Wang:2022neq,Wang:2022gfb,Wang:2022ltr,Wang:2021itn}. For more investigation with the assumption of meson-baryon molecular state, we refer to the Refs.~\cite{Chen:2015loa,Chen:2015moa,He:2015cea,Meissner:2015mza,Roca:2015dva,Chen:2020opr,Yan:2021nio,Wu:2021caw,Chen:2020uif,Phumphan:2021tta,Du:2021fmf,Lu:2021irg,Chen:2021obo,Xiao:2019gjd,Wang:2019nvm,Xiao:2019gjd}. Though the pentaquark states were investigated deeply via the different models as mentioned above, there still exists ambiguity in their  properties, and they need further investigation to be identified clearly. Such investigations may either provide insights into the observed states about their properties or some of them may focus on their decay modes other than the observation channels. Moreover, some new candidate states can also be offered from the analyses for different possible pentaquarks' spectroscopic properties or decay channels. One can find such investigations in Refs.~\cite{Feijoo:2015kts,Liu:2021ixf,Chen:2015sxa,Wang:2015jsa,Kubarovsky:2015aaa,Karliner:2015voa,Cheng:2015cca,Liu:2020ajv,Yang:2021pio,Stancu:2021rro,Dong:2020nwk,Voloshin:2019aut,Ling:2021lmq,Xing:2021yid,Cheng:2021gca,Wang:2019hyc,Xu:2019zme,Liu:2021ojf}.

As is seen from all these researches, the pentaquark investigations are among the very hot topics, not only including the survey of the observed related states but also of the ones that have the potentials to be observed in future experiments. Ongoing improvements in both experimental facilities and techniques make the expectation for the observations of such new states natural. In this respect, there occurred many theoretical efforts to offer such new states to shed light on experimental investigations. With this motivation, in Refs.~\cite{Chen:2016heh,Liu:2017xzo,Azizi:2017bgs,Huang:2021ave,Zhu:2020vto,Ferretti:2018ojb,Shimizu:2016rrd,Yang:2022bfu,Liu:2021efc,Liu:2021tpq,Xie:2020ckr,Zhang:2020dwp,Paryev:2020jkp,Xie:2020niw,Ferretti:2020ewe,Cao:2019gqo,Wang:2019zaw,Gutsche:2019mkg,Wang:2019ato,Huang:2018wed,Yang:2018oqd,Yamaguchi:2017zmn,Zhang:2020cdi} the pentaquark states involving bottom quark were investigated. Pentaquarks with full heavy quark content were studied in Refs.~\cite{Wang:2021xao,An:2020jix,Yan:2021glh,Zhang:2020vpz}. The expectation of new pentaquark states with quark content other than the first observed states has been advocated by the observation of new states with the strange quark, that is, $P_{cs}(4459)^0$ and $P_{cs}(4337)^+$. Inspired by the observed pentaquark states with and without strange quark, the observed states with strange quark were analyzed~\cite{Azizi:2021utt,Zhu:2021lhd,Cheng:2021gca,Yan:2021nio,Yang:2021pio,Wang:2021itn,Wu:2021caw,Chen:2020uif,Liu:2020ajv,Wang:2020eep,Peng:2020hql,Chen:2020kco,Chen:2021tip,Wang:2022gfb,Xiao:2021rgp,Ozdem:2021ugy,Clymton:2021thh,Du:2021bgb,Giron:2021fnl,Ferretti:2021zis,Wang:2022mxy,Shi:2021wyt,Deng:2022vkv,Yan:2022wuz,Hu:2021nvs}. Besides the observed strange pentaquark states, the possible states containing double or triple strange quarks were taken into account in the Refs.~\cite{Wang:2021hql,Wang:2015wsa,Azizi:2021pbh,Wang:2022neq,Shi:2021wyt,Deng:2022vkv,Yan:2022wuz,Wang:2020bjt,Azizi:2018dva,Meng:2019fan,Ozdem:2022iqk} to provide insights for the future experiments by supplying information such as their possible spectroscopic properties or decay mechanisms.  This natural expectation of the forthcoming observations of the new pentaquark states with possible different substructures has also motivated  us to study the properties of such possible new states. With this motivation, in Ref~\cite{Azizi:2021pbh}, we analyzed the possible double strange pentaquark and calculated the probable mass and decay width for this state. In this work, we aim to make a similar analysis for a candidate state of triple strange pentaquark by investigating the decay width for the strong $P_{csss}\rightarrow\Omega^-J/\psi$ transition by applying the QCD sum rule method~\cite{Shifman:1978bx,Shifman:1978by,Ioffe81}. This method has yielded many successful predictions consistent with the experimental findings up to now, which puts it among the powerful nonperturbative approaches. In this method, the main ingredient is a proper interpolating current, and in this work, we take into account two of them, which were also suggested in Ref.~\cite{Wang:2015wsa}. In that work, the masses for the pentaquark states were predicted with these interpolating fields, and we apply these mass values as inputs in our analyses. These currents are in diquark-diquark-antiquark form with spin-parity quantum numbers $J^P=\frac{1}{2}^-$ and can couple to both the negative and positive parity states. Therefore in the analyses, we consider the possible pentaquark candidate states with both negative and positive parities that couple to each interpolating current and obtain their corresponding decay widths.   

The present article contains the following outline: In Section~\ref{II} we give the details of the QCD sum rule calculations to obtain the QCD sum rules for the coupling constants entering the decays calculated for both the currents used for the triple pentaquark state, which will be represented as $P_{csss}$ in the text. Section~\ref{III} presents the numerical analyses for the obtained QCD sum rules. The last section contains the  summary and conclusion.

\section{The strong decay $P_{csss} \rightarrow \Omega^-J/\psi $}\label{II}

 In this section, we provide the analyses for the strong decay of candidate pentaquark state, $P_{csss} \rightarrow \Omega^-J/\psi $, to obtain the coupling constants. These coupling constants are among the main ingredients of the related decay width calculations. To get the  coupling constants via the QCD sum rule method the three-point correlation function given below is used:  
\begin{equation}
\Pi_{\mu\mu'} (p, q)=i^2\int d^{4}xe^{-ip\cdot
x}\int d^{4}ye^{ip'\cdot y}\langle 0|\mathcal{T} \{J_{\mu}^{\Omega^-}(y)
J_{\mu'}^{J/\psi}(0)\bar{J}^{P_{csss}}(x)\}|0\rangle,
\label{eq:CorrF1Pc}
\end{equation}
where $J^{P_{csss}}$ is the interpolating current for the candidate $P_{csss}$ pentaquark. In this work we consider two different possible currents with the following forms:
%
\begin{eqnarray}
J_{P_{csss}}&=&\epsilon^{ila}\epsilon^{ijk}\epsilon^{lmn}s^{T}_{j}C\gamma_{\alpha} s_k s^{T}_m C\gamma^{\alpha} c_{n} C \bar{c}^{T}_a\label{CurrentPcsss1},\nonumber\\
\end{eqnarray}
and
\begin{eqnarray}
J_{P_{csss}}&=&\frac{1}{\sqrt{3}}\epsilon^{ila}\epsilon^{ijk}\epsilon^{lmn}s^{T}_{j}C\gamma_{\alpha} s_k s^{T}_m C\gamma_{5} c_{n} \gamma_5 \gamma^{\alpha} C \bar{c}^{T}_a \label{CurrentPcsss2}.\nonumber\\
\end{eqnarray}
As is seen in Eq.~(\ref{eq:CorrF1Pc}), besides the current of the pentaquark state, one also needs the currents for the $\Omega^-$ and $J/\psi$ final states, which are  
\begin{eqnarray}
J_{\mu}^{\Omega^-}&=&\epsilon^{lmn}(s^{T}_l C\gamma_{\mu} s_m) s_n,\nonumber\\
J_{\mu'}^{J/\psi}&=&\bar{c}_l\gamma_{\mu'}c_l.
\label{InterpFields}
\end{eqnarray}
In  Eqs.~(\ref{CurrentPcsss1}), (\ref{CurrentPcsss2}) and (\ref{InterpFields}) the subindices, $i,~j,~k,~a,~l,~m,~n$, represent the color indices, and $s,~c$ are used to symbolize the corresponding quark fields and $C$ is the charge conjugation operator. The calculation of the correlation function requires some standard steps separated into two parts. In the first part, a hadronic representation of it is obtained, in the other part, a representation is attained in terms of QCD degrees of freedom. A proper match of both parts via a dispersion relation results in the physical parameters obtained in terms of QCD ones. In this work, these parameters are the coupling constants, $g_1$, $g_2$ and $g_3$ for the transition of pentaquark state with negative parity and $f_1$, $f_2$, and $f_3$ for the transition of pentaquark state with positive parity that are obtained in terms of perturbative and nonperturbative QCD degrees of freedom as well as some auxiliary parameters involved. 

To calculate the correlator's hadronic representation in terms of hadronic degrees of freedom, we insert complete sets of hadronic states, which have same quantum numbers with the considered interpolating currents, into the correlation function. After taking the four integrals, we get the results of this side as
\begin{eqnarray}
\Pi _{\mu\mu' }^{\mathrm{Had}}(p, q)&=&\frac{\langle 0|J_{\mu}^{\Omega^- }|\Omega^-(p',s')\rangle \langle 0|J_{\mu' }^{J/\psi}|J/\psi(q)\rangle \langle J/\psi(q) \Omega^-(p',s')|P_{csss}(p,s)\rangle \langle P_{csss}(p,s)|\bar{J}^{P_{csss}}|0\rangle }{(m_{\Omega^-}^2-p'^2)(m_{J/\psi}^2-q^2)(m_{P_{csss}}^2-p^2)}\nonumber\\
&+&\frac{\langle 0|J_{\mu}^{\Omega^- }|\Omega^-(p',s')\rangle \langle 0|J_{\mu' }^{J/\psi}|J/\psi(q)\rangle \langle J/\psi(q) \Omega^-(p',s')|\widetilde{P}_{csss}(p,s)\rangle \langle \widetilde{P}_{csss}(p,s)|\bar{J}^{P_{csss}}|0\rangle }{(m_{\Omega^-}^2-p'^2)(m_{J/\psi}^2-q^2)(m_{\widetilde{P}_{csss}}^2-p^2)}+\cdots.  \label{eq:CorrF1Phys}
\end{eqnarray}  
In Eq.~(\ref{eq:CorrF1Phys}), to represent negative and positive parity one-particle pentaquark states, we respectively  use $|P_{csss}(p,s)\rangle$ and $|\widetilde{P}_{csss}(p,s)\rangle$;  and $m_{P_{csss}}$ and $m_{\widetilde{P}_{csss}}$ are their corresponding masses. Similarly, $|\Omega^-(p',s')\rangle$, $|J/\psi(q)\rangle $, $m_{\Omega^-}$ and $m_{J/\psi}$ are the one-particle states and masses for the $\Omega^-$ and $J/\psi$ states, respectively. The matrix elements between vacuum and one-particle states present in Eq.~(\ref{eq:CorrF1Phys}) are defined in terms of the current coupling constants, spinors and masses as
\begin{eqnarray}
\langle 0|J^{P_{csss}}|P_{csss}(p,s)\rangle &=&\lambda_{P_{csss}} u_{P_{css}}(p,s),
\nonumber\\
\langle 0|J^{P_{csss}}|\widetilde{P}_{csss}(p,s)\rangle &=&\lambda_{\widetilde{P}_{csss}}\gamma_5 u_{\widetilde{P}_{css}}(p,s),
\nonumber\\
\langle 0|J_{\mu}^{\Omega^- }|\Omega^-(p',s')\rangle &=&\lambda_{\Omega^-} u_{\Omega^-,\mu}(p',s'),
\nonumber\\
\langle 0|\eta_{\mu' }^{J/\psi}|J/\psi(q)\rangle &=&f_{J/\psi} m_{J/\psi} \varepsilon_{\mu'},
\label{eq:MatPcsss}
\end{eqnarray}
where $\varepsilon_{\mu'}$ and  $ f_{J/\psi} $ are polarization vector and decay constant of $J/\psi$ and $\lambda_{P_{csss}}$, $\lambda_{\widetilde{P}_{csss}}$ and $\lambda_{\Omega^-}$ are current coupling constants of the negative, positive parity pentaquark states and $\Omega^-$ state, respectively. The masses $m_{P_{csss}}$ and $m_{\widetilde{P}_{csss}}$  and the current coupling constants  $\lambda_{P_{csss}}$ and $\lambda_{\widetilde{P}_{csss}}$ are calculated in Ref.~\cite{Wang:2015wsa} using the same currents as  the present study. We will use them as inputs in the analyses of the considered strong decays. The spinors for the states are represented as $ u_{P_{csss}}$,  $ u_{\widetilde{P}_{csss}}$ and $ u_{\Omega^-,\mu}$ which satisfy
\begin{eqnarray}
\sum_{s}u_{P_{csss}}(p,s)\bar{u}_{P_{csss}}(p,s)&=&({\slashed
p}+m_{P_{csss}}),\nonumber \\
\sum_{s}u_{P_{csss}}(p,s)\bar{u}_{\widetilde{P}_{csss}}(p,s)&=&({\slashed
p}+m_{\widetilde{P}_{csss}}),\nonumber \\
\sum_{s'}u_{\Omega^-,\mu}(p',s')\bar{u}_{\Omega^-,\nu}(p',s')&=&-({\slashed
p'}+m_{\Omega^-})\Big[g_{\mu\nu}-\frac{1}{3}\gamma_{\mu}\gamma_{\nu}-\frac{2p'_{\mu}p'_{\nu}}{3m_{\Omega^-}^2}+\frac{p'_{\mu}\gamma_{\nu}-p'_{\nu}\gamma_{\mu}}{3m_{\Omega^-}}\Big], \nonumber\\
\varepsilon_{\alpha}\varepsilon^*_{\beta}&=&-g_{\alpha\beta}+\frac{q_\alpha q_\beta}{m_{J/\psi}^2}.
\label{eq:SumPc}
\end{eqnarray}
The matrix elements, $\langle J/\psi(q) \Omega^-(p',s')|P_{csss}(p,s)\rangle$ and $\langle J/\psi(q) \Omega^-(p',s')|\widetilde{P}_{csss}(p,s)\rangle$ have the following forms in terms of coupling constants of the interested interactions
\begin{eqnarray}
\langle J/\psi(q) \Omega^-(p',s')|P_{csss}(p,s)\rangle &=&\bar{u}_{\Omega,\alpha}(p',s')\Bigg\{g_1(q_{\alpha}\not\!\varepsilon -\varepsilon_{\alpha}\not\!q)+g_2(P.\varepsilon q_{\alpha}-P.q\varepsilon_{\alpha})+g_3(q.\varepsilon q_{\alpha}-q^2\varepsilon_{\alpha})\Bigg\}u_{P_{csss}(p,s)},\nonumber\\
\langle J/\psi(q) \Omega^-(p',s')|\widetilde{P}_{csss}(p,s)\rangle& = & \bar{u}_{\Omega,\alpha}(p',s')\Bigg\{f_1(q_{\alpha}\not\!\varepsilon -\varepsilon_{\alpha}\not\!q)+f_2(P.\varepsilon q_{\alpha}-P.q\varepsilon_{\alpha})+f_3(q.\varepsilon q_{\alpha}-q^2\varepsilon_{\alpha})\Bigg\}\gamma_5 u_{\widetilde{P}_{csss}(p,s)},\nonumber\\
\label{eq:Matvertex}
\end{eqnarray}
where $P=(p+p')/2$. Insertion of all these matrix elements given in Eqs.~(\ref{eq:MatPcsss}) and (\ref{eq:Matvertex}) into Eq.~(\ref{eq:CorrF1Phys}) and using the Eq.~(\ref{eq:SumPc}) to make the summations over the spinors and polarization vectors, gives us the final form of the hadronic side after which double Borel transformations with respect to $p^2$ and $p'^2$ are applied.  The Borel transformations supply a suppression over the contributions coming from higher states and continuum. The result contains many Lorentz structures, and the coefficients of some of them will be used in the analyses to obtain the coupling constants. We will take these coefficients, which belong to the same Lorentz structure, from the hadronic side and QCD side and equate them. By employing this procedure, we get some coupled equations to be solved for the coupling constants. In the below equation, we represent the correlator for the hadronic side in terms of these Lorentz structures keeping only the ones we directly use in our analyses. 
\begin{eqnarray}
\tilde{\Pi}_{\mu\mu' }^{\mathrm{Had}}(p, q)&=&-e^{-\frac{m_{P_{csss}^2}}{M^2}}e^{-\frac{m_{\Omega^-}^2}{M'^2}}\frac{f_{J/\psi} \lambda_{\Omega^-} \lambda_{P_{csss}} m_{J/\psi}}{(m_{J/\psi}^2 + Q^2)}\Bigg[ -\frac{1}{3}\Big(g_2 m_{\Omega^-} (m_{\Omega^-}^2 - m_{P_{csss}}^2) - 2 g_3 m_{\Omega^-} Q^2 + 
 g_1 (m_{\Omega^-} m_{P_{csss}} \nonumber\\
 &+& 2 m_{P_{csss}}^2 + 2 Q^2)\Big)p_{\mu}\gamma_{\mu'} + \frac{1}{6 m_{\Omega^-}}\Big(-4 g_1 (m_{\Omega^-} - m_{P_{csss}}) +
 2 g_3 (m_{\Omega^-}^2 - m_{\Omega^-} m_{P_{csss}} + m_{P_{csss}}^2 - Q^2) \nonumber\\
 & + &
 g_2 (3 m_{\Omega^-}^2 - m_{\Omega^-} m_{P_{csss}} - m_{P_{csss}}^2 + Q^2)\Big)\not\!p'p_{\mu}p'_{\mu'}+\frac{1}{6}\Big(-4 g_1 m_{\Omega^-}- (g_2 - 2 g_3) [2 m_{\Omega^-}^2 + m_{\Omega^-} m_{{P_{csss}}} \nonumber\\
 &+& 
   2 (m_{P_{csss}}^2 + Q^2)]\Big)p_{\mu} p_{\mu'} -\frac{1}{6}\Big(-4 g_1 (m_{\Omega^-} + m_{P_{csss}}) + 
 g_2 (m_{\Omega^-} m_{P_{csss}} + 4 m_{P_{csss}}^2 + 2 Q^2) + 
 2 g_3 (2 m_{\Omega^-}^2 \nonumber\\
 &+&
  m_{\Omega^-} m_{P_{csss}} + 2 m_{P_{csss}}^2 + 4 Q^2)\Big)p_{\mu} p'_{\mu'}-\frac{m_{\Omega^-}}{2}\Big(-2 g_1 m_{\Omega^-} + g_2 m_{\Omega^-}^2 + 2 g_1 m_{P_{csss}} - g_2 m_{P_{csss}}^2\nonumber\\
  & -& 
 2 g_3 Q^2 \Big)\not\!p g_{\mu\mu'}+\frac{1}{2}\Big((m_{\Omega^-}^2 + m_{\Omega^-} m_{P_{csss}} + m_{P_{csss}}^2 + 
   Q^2) (2 g_1 (m_{\Omega^-} - m_{P_{csss}}) + g_2 (-m_{\Omega^-}^2 + m_{P_{csss}}^2)\nonumber\\
   & +& 
   2 g_3 Q^2)\Big)g_{\mu\mu'}+\mathrm{other~structures}\Bigg] 
 \nonumber\\
&-&e^{-\frac{m_{\widetilde{P}_{csss}}^2}{M^2}}e^{-\frac{m_{\Omega^-}^2}{M'^2}}\frac{f_{J/\psi} \lambda_{\Omega^-} \lambda_{\widetilde{P}_{csss}}m_{J/\psi}}{(m_{J/\psi}^2 + Q^2)}\Bigg[ \frac{1}{3}\Big(f_2 m_{\Omega^-} (m_{\Omega^-}^2 - m_{\widetilde{P}_{csss}}^2) - 2 f_3 m_{\Omega^-} Q^2 + 
 f_1 (-m_{\Omega^-} m_{\widetilde{P}_{csss}} \nonumber\\
 &+& 2 m_{\widetilde{P}_{csss}}^2 + 2 Q^2)\Big)p_{\mu}\gamma_{\mu'}+ \frac{1}{6 m_{\Omega^-}}\Big(4 f_1 (m_{\Omega^-} + m_{\widetilde{P}_{csss}}) -
 2 f_3 (m_{\Omega^-}^2 + m_{\Omega^-} m_{\widetilde{P}_{csss}} + m_{\widetilde{P}_{csss}}^2 - Q^2) \nonumber\\
 & - &
 f_2 (3 m_{\Omega^-}^2 + m_{\Omega^-} m_{\widetilde{P}_{csss}} - m_{\widetilde{P}_{csss}}^2 + Q^2)\Big)\not\!p'p_{\mu}p'_{\mu'}+\frac{1}{6}\Big(4 f_1 m_{\Omega^-}+ (f_2 - 2 f_3) [2 m_{\Omega^-}^2 - m_{\Omega^-} m_{\widetilde{P}_{csss}} \nonumber\\
 &+& 
   2 (m_{\widetilde{P}_{csss}}^2 + Q^2)]\Big)p_{\mu} p_{\mu'}+\frac{1}{6}\Big(-4 f_1 (m_{\Omega^-} - m_{\widetilde{P}_{csss}}) + 
 f_2 (-m_{\Omega^-} m_{\widetilde{P}_{csss}} + 4 m_{\widetilde{P}_{csss}}^2 + 2 Q^2) + 
 f_3 (4 m_{\Omega^-}^2 \nonumber\\
 &-&
  2 m_{\Omega^-} m_{\widetilde{P}_{csss}} + 4 m_{\widetilde{P}_{csss}}^2 + 8 Q^2)\Big)p_{\mu} p'_{\mu'}-\frac{m_{\Omega^-}}{2}\Big(-f_2 m_{\Omega^-}^2 + f_2 m_{\widetilde{P}_{csss}}^2 + 2 f_1 (m_{\Omega^-} + m_{\widetilde{P}_{csss}}) \nonumber\\
  &+& 2 f_3 Q^2\Big)\not\!p g_{\mu\mu'}-\frac{1}{2}\Big( (m_{\Omega^-}^2 - m_{\Omega^-} m_{\widetilde{P}_{csss}} + m_{\widetilde{P}_{csss}}^2 + 
   Q^2) (2 f_1 (m_{\Omega^-} + m_{\widetilde{P}_{csss}}) + f_2 (-m_{\Omega^-}^2 + m_{\widetilde{P}_{csss}}^2) \nonumber\\
&+& 
2 f_3 Q^2)\Big)g_{\mu\mu'}+ \mathrm{other~structures}
  \Bigg] +\cdots,
\label{eq:had1}
\end{eqnarray}
where $Q^2=-q^2$; and  $M^2$ and $M'^2$ are the Borel parameters. The "other structures" in Eq.~(\ref{eq:had1}) represents  the other present structures that are not used in the analyses, and $\cdots$ is used to indicate the contributions of higher states and continuum. As is seen from Eq.~(\ref{eq:had1}), we have six coupling constants, to obtain them we need six equations, and we get these six equations from coefficients of six different Lorentz structures obtained in both the hadronic and QCD sides.

\begin{figure}[h!]
\begin{center}
\includegraphics[totalheight=10cm,width=10cm]{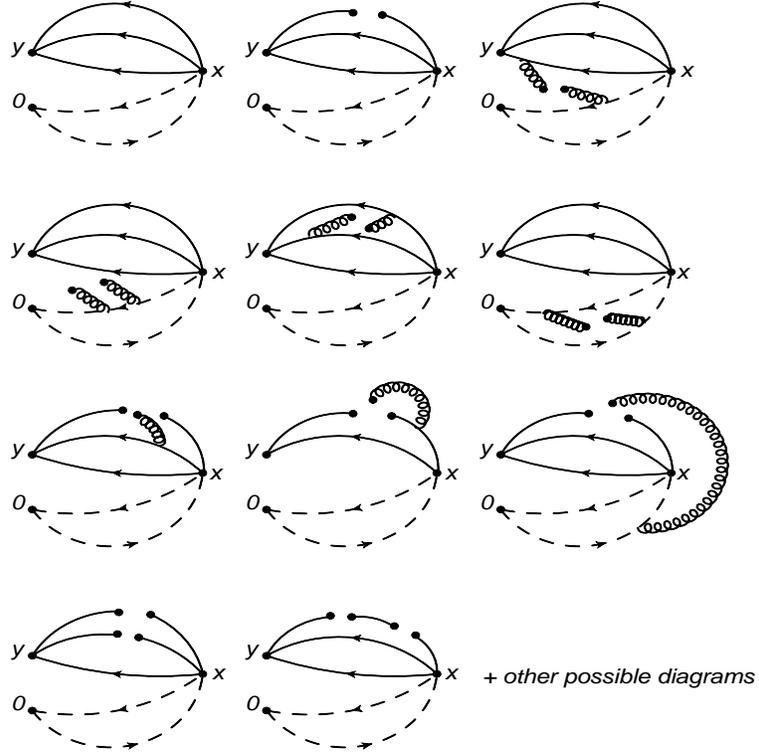}
\end{center}
\caption{The Feynman diagrams corresponding to the terms considered in the QCD side of the strong decay calculations. The light   and  heavy quarks  are represented by the solid and dashed lines, respectively.}
\label{FeynmanDiagram}
\end{figure}

As stated, the coupling constants are obtained from the solutions of the coupled equations that are constructed from the matching of the coefficients of the same Lorentz structures calculated in both the  hadronic and QCD sides. Hence, we  need  to have the results for the QCD side as well.  To this end, we calculate the correlation function, Eq.~(\ref{eq:CorrF1Pc}), in quark-gluon language and in terms of the  fundamental QCD degrees of freedom in deep Euclidean region. To proceed, the explicit forms of the interpolating currents given in Eqs.~(\ref{CurrentPcsss1})/ (\ref{CurrentPcsss2}) and (\ref{InterpFields}) are inserted into the correlation function. Using the Wick's theorem, we contract all  the  quark fields giving the results in terms of the light and heavy quark propagators. In this step, we find
\begin{eqnarray}
\Pi_{\mu\mu' }^\mathrm{OPE}(p,p',q)&=&i^2\int d^{4}xe^{-ip\cdot x}\int d^{4}ye^{ip'\cdot y}\epsilon^{klm}\epsilon^{i'l'a'}\epsilon^{i'j'k'}\epsilon^{l'm'n'} 2 \bigg\{Tr[\gamma^{\nu}CS_{s}^{Tkj'}(y-x)C \gamma_{\mu} S_{s}^{lk'}(y-x)]\nonumber\\
&\times&  S_s^{mm'}(y-x) \gamma_{\nu}C S_{c}^{Tnn'}(-x) C \gamma_{\mu'} C S_{c}^{Ta'n}(x)C +
S_{s}^{mj'}(y-x)\gamma^{\nu}CS_{s}^{Tkk'}(y-x) C \gamma_{\mu} S_{s}^{lm'}(y-x)\nonumber\\
&\times& 
\gamma_{\nu}C S_{c}^{Tnn'}(-x) C \gamma_{\mu'}C S_{c}^{Ta'n}(x)C
+S_{s}^{mk'}(y-x)\gamma^{\nu}CS_{s}^{Tlj'}(y-x) C \gamma_{\mu} S_{s}^{km'}(y-x)\nonumber\\
&\times& 
\gamma_{\nu}C S_{c}^{Tnn'}(-x) C \gamma_{\mu'}C S_{c}^{Ta'n}(x)C\bigg\},\label{eq:CorrF1Theore1C1}
\end{eqnarray}
for the first current, given in Eq.~(\ref{CurrentPcsss1}) and
\begin{eqnarray}
\Pi_{\mu\mu' }^\mathrm{OPE}(p,p',q)&=&\frac{-i^2}{\sqrt{3}}\int d^{4}xe^{-ip\cdot x}\int d^{4}ye^{ip'\cdot y}\epsilon^{klm}\epsilon^{i'l'a'}\epsilon^{i'j'k'}\epsilon^{l'm'n'} 2 \bigg\{Tr[\gamma^{\nu}CS_{s}^{Tkj'}(y-x)C \gamma_{\mu} S_{s}^{lk'}(y-x)]\nonumber\\
&\times&  S_s^{mm'}(y-x) \gamma_{5}C S_{c}^{Tnn'}(-x) C \gamma_{\mu'} C S_{c}^{Ta'n}(x)C\gamma_{\nu}\gamma_{5} +
S_{s}^{mj'}(y-x)\gamma^{\nu}CS_{s}^{Tkk'}(y-x) C \gamma_{\mu} S_{s}^{lm'}(y-x)\nonumber\\
&\times& 
\gamma_{5}C S_{c}^{Tnn'}(-x) C \gamma_{\mu'}C S_{c}^{Ta'n}(x)C\gamma_{\nu}\gamma_{5}
+S_{s}^{mk'}(y-x)\gamma^{\nu}CS_{s}^{Tlj'}(y-x) C \gamma_{\mu} S_{s}^{km'}(y-x)\nonumber\\
&\times& 
\gamma_{5}C S_{c}^{Tnn'}(-x) C \gamma_{\mu'}C S_{c}^{Ta'n}(x)C\gamma_{\nu}\gamma_{5}\bigg\},\label{eq:CorrF1Theore1C2}
\end{eqnarray}
for the second current, given in Eq.~(\ref{CurrentPcsss2}).
As it is seen from the above equations, to proceed we need to have the explicit forms of the quark propagators in $ x $-space. We use the following  light and heavy quark propagators~\cite{Yang:1993bp,Reinders:1984sr} :
\begin{eqnarray}
S_{s,}{}_{ab}(x)&=&i\delta _{ab}\frac{\slashed x}{2\pi ^{2}x^{4}}-\delta _{ab}%
\frac{m_{s}}{4\pi ^{2}x^{2}}-\delta _{ab}\frac{\langle \overline{s}s\rangle
}{12} +i\delta _{ab}\frac{\slashed xm_{s}\langle \overline{s}s\rangle }{48}%
-\delta _{ab}\frac{x^{2}}{192}\langle \overline{s}g_{\mathrm{s}}\sigma
Gs\rangle +i\delta _{ab}\frac{x^{2}\slashed xm_{s}}{1152}\langle \overline{s}%
g_{\mathrm{s}}\sigma Gs\rangle  \notag \\
&&-i\frac{g_{\mathrm{s}}G_{ab}^{\alpha \beta }}{32\pi ^{2}x^{2}}\left[ %
\slashed x{\sigma _{\alpha \beta }+\sigma _{\alpha \beta }}\slashed x\right]
-i\delta _{ab}\frac{x^{2}\slashed xg_{\mathrm{s}}^{2}\langle \overline{s}%
s\rangle ^{2}}{7776} ,  \label{Eq:qprop}
\end{eqnarray}%
and
\begin{eqnarray}
S_{c,{ab}}(x)&=&\frac{i}{(2\pi)^4}\int d^4k e^{-ik \cdot x} \left\{
\frac{\delta_{ab}}{\!\not\!{k}-m_c}
-\frac{g_sG^{\alpha\beta}_{ab}}{4}\frac{\sigma_{\alpha\beta}(\!\not\!{k}+m_c)+
(\!\not\!{k}+m_c)\sigma_{\alpha\beta}}{(k^2-m_c^2)^2}\right.\nonumber\\
&&\left.+\frac{\pi^2}{3} \langle \frac{\alpha_sGG}{\pi}\rangle
\delta_{ij}m_c \frac{k^2+m_c\!\not\!{k}}{(k^2-m_c^2)^4}
+\cdots\right\},
 \label{Eq:Qprop}
\end{eqnarray}
where $G^{\alpha\beta}_{ab}=G^{\alpha\beta}_{A}t_{ab}^{A}$ and $GG=G_{A}^{\alpha\beta}G_{A}^{\alpha\beta}$ with $a,~b=1,2,3$ and $A=1,2,\cdots,8$.  Here, $t^A=\frac{\lambda^A}{2}$ with  $\lambda^A$ being the  Gell-Mann matrices.  After using the propagators in Eqs. (\ref{eq:CorrF1Theore1C1}) and (\ref{eq:CorrF1Theore1C2}), we apply the Fourier transformation by performing the  integral  over $x$   to transform the results to the  momentum space. 
In terms of the Feynman diagrams, the calculations done up to this point are equivalent to the calculations of the  Feynman diagrams shown in figure~\ref{FeynmanDiagram} directly in the momentum space. We shall say that our calculations cover all the possible diagrams up to dimension six nonperturbative operators, part of which is shown explicitly in figure~\ref{FeynmanDiagram}.

The procedures above lead to the following expression for the QCD side in terms of different Lorentz structures:
\begin{eqnarray}
\Pi_{\mu\mu',a}^{OPE}(p,q)&=&\Pi_{a,1}\,p_{\mu}\gamma_{\mu'}+\Pi_{a,2}\,\not\!p'p_{\mu}p'_{\mu'}+
\Pi_{a,3}\,p_{\mu} p_{\mu'}+\Pi_{a,4}\,p_{\mu} p'_{\mu'}+\Pi_{a,5}\,\not\!p g_{\mu\mu'}+\Pi_{a,6}\,g_{\mu\mu'}+\mathrm{other \,\,\, structures},\notag \\ \label{eq:PiOPE}
\end{eqnarray}
with ``other structures" representing the contributions of  other possible structures.  We select the coefficients of the explicitly shown structures in this equation in the QCD side. The sub-index $a=1,~2$ in  $\Pi_{\mu\mu',a}^{OPE}$ is used  to indicate the result for either the first pentaquark current or the second one. From the calculations we obtain the spectral densities defined as $\rho_{a,i}(s,s',q^2)=\frac{1}{\pi}Im(\Pi_{a,i})$, with subindices $i=1,~2,~\cdots,6$ corresponding to the considered Lorentz structures. These spectral densities are used in the following dispersion relation:
\begin{eqnarray}
\Pi_{a,i}=\int ds\int
ds'\frac{\rho_{a,i}^{\mathrm{pert}}(s,s',q^{2})+\rho_{a,i}^{\mathrm{non-pert}}(s,s',q^{2})}{(s-p^{2})
(s'-p'^{2})}, \label{eq:Pispect}
\end{eqnarray}
where we represent the spectral densities obtained for the perturbative part with $\rho_{a,i}^{\mathrm{pert}}(s,s',q^{2})$ and for non-perturbative part with $\rho_{a,i}^{\mathrm{non-pert}}(s,s',q^{2})$, expressions of which are very lengthy and we do not present them here. The next step is to apply the Borel transformation with respect to the variables $ p^2 $ and $ p'^2 $ and perform the continuum subtraction supplied by the quark-hadron duality assumption. These procedures bring two continuum threshold parameters $ s_0 $ and $ s'_0 $ respectively in the initial and final state besides the two Borel parameters previously discussed.

The matches of the coefficients of the given Lorentz structures obtained from both sides provide us with six coupled equations to be solved for the required strong coupling constants:
\begin{eqnarray}
&&A\frac{1}{3}\Big(g_2 m_{\Omega^-} (m_{\Omega^-}^2 - m_{P_{csss}}^2) - 2 g_3 m_{\Omega^-} Q^2 + 
 g_1 (-m_{\Omega^-} m_{P_{csss}} + 2 m_{P_{csss}}^2 + 2 Q^2)\Big)-B\frac{1}{3}\Big(f_2 m_{\Omega^-} (m_{\Omega^-}^2 - m_{\widetilde{P}_{csss}}^2) \nonumber\\
 &&- 2 f_3 m_{\Omega^-} Q^2 + 
 f_1 (-m_{\Omega^-} m_{\widetilde{P}_{csss}} + 2 m_{\widetilde{P}_{csss}}^2 + 2 Q^2)=\tilde{\Pi}_{a,1},\nonumber\\
 &&-A\frac{1}{6 m_{\Omega^-}}\Big(-4 g_1 (m_{\Omega^-} - m_{P_{csss}}) +
 2 g_3 (m_{\Omega^-}^2 - m_{\Omega^-} m_{P_{csss}} + m_{P_{csss}}^2 - Q^2) 
  + 
 g_2 (3 m_{\Omega^-}^2 - m_{\Omega^-} m_{P_{csss}} - m_{P_{csss}}^2\nonumber\\
 &&+ Q^2)\Big)-B\frac{1}{6 m_{\Omega^-}}\Big(4 f_1 (m_{\Omega^-} + m_{\widetilde{P}_{csss}}) -
 2 f_3 (m_{\Omega^-}^2 + m_{\Omega^-} m_{\widetilde{P}_{csss}} + m_{\widetilde{P}_{csss}}^2 - Q^2) 
  - 
 f_2 (3 m_{\Omega^-}^2 + m_{\Omega^-} m_{\widetilde{P}_{csss}} - m_{\widetilde{P}_{csss}}^2 \nonumber\\
 &&+ Q^2)\Big)=\tilde{\Pi}_{a,2},\nonumber\\
 &&-A\frac{1}{6}\Big(-4 g_1 m_{\Omega^-}- (g_2 - 2 g_3) [2 m_{\Omega^-}^2 + m_{\Omega^-} m_{P_{csss}} + 
   2 (m_{P_{csss}}^2 + Q^2)]\Big)-B\frac{1}{6}\Big(4 f_1 m_{\Omega^-}+ (f_2 - 2 f_3) [2 m_{\Omega^-}^2  \nonumber\\
 &&- m_{\Omega^-} m_{\widetilde{P}_{csss}}+ 
   2 (m_{\widetilde{P}_{csss}}^2 + Q^2)]\Big)=\tilde{\Pi}_{a,3},\nonumber\\
   &&-A\frac{1}{6}\Big(-4 f_1 (m_{\Omega^-} + m_{P_{csss}}) + 
 g_2 (m_{\Omega^-} m_{P_{csss}} + 4 m_{P_{csss}}^2 + 2 Q^2) + 
 2 g_3 (2 m_{\Omega^-}^2 +
  m_{\Omega^-} m_{P_{csss}} + 2 m_{P_{csss}}^2 + 4 Q^2)\Big)\nonumber\\
  &&-B\frac{1}{6}\Big(-4 f_1 (m_{\Omega^-} - m_{\widetilde{P}_{csss}}) + 
 f_2 (-m_{\Omega^-} m_{\widetilde{P}_{csss}} + 4 m_{\widetilde{P}_{csss}}^2 + 2 Q^2) + 
 f_3 (4 m_{\Omega^-}^2 -
  2 m_{\Omega^-} m_{\widetilde{P}_{csss}} + 4 m_{\widetilde{P}_{csss}}^2 + 8 Q^2)\Big)=\tilde{\Pi}_{a,4},\nonumber\\
 &&A\frac{m_{\Omega^-}}{2}\Big(-2 g_1 m_{\Omega^-} + g_2 m_{\Omega^-}^2 + 2 g_1 m_{P_{csss}} - g_2 m_{P_{csss}}^2 - 
 2 g_3 Q^2 \Big)+B\frac{m_{\Omega^-}}{2}\Big(-f_2 m_{\Omega^-}^2 + f_2 m_{\widetilde{P}_{csss}}^2 + 2 f_1 (m_{\Omega^-} + m_{\widetilde{P}_{csss}}) \nonumber\\
  &&+ 2 f_3 Q^2\Big) =\tilde{\Pi}_{a,5},\nonumber\\
 &&-A\frac{1}{2}\Big((m_{\Omega^-}^2 + m_{\Omega^-} m_{P_{csss}} + m_{P_{csss}}^2 + 
   Q^2) (2 g_1 (m_{\Omega^-} - m_{P_{csss}}) + g_2 (-m_{\Omega^-}^2 + m_{P_{csss}}^2)+ 
   2 g_3 Q^2)\Big)-B\frac{1}{2}\Big( (m_{\Omega^-}^2 \nonumber\\
  &&- m_{\Omega^-} m_{\widetilde{P}_{csss}} + m_{\widetilde{P}_{csss}}^2 + 
   Q^2) (2 f_1 (m_{\Omega^-} + m_{\widetilde{P}_{csss}}) + f_2 (-m_{\Omega^-}^2 + m_{\widetilde{P}_{csss}}^2) + 
2 f_3 Q^2)\Big)=\tilde{\Pi}_{a,6},
\end{eqnarray}
where
\begin{eqnarray}
A&=&e^{-\frac{m_{P_{csss}^2}}{M^2}}e^{-\frac{m_{\Omega^-}^2}{M'^2}}\frac{f_{J/\psi} \lambda_{\Omega^-} \lambda_{P_{csss}} m_{J/\psi}}{(m_{J/\psi}^2 + Q^2)},\nonumber\\
B&=&e^{-\frac{m_{\widetilde{P}_{csss}}^2}{M^2}}e^{-\frac{m_{\Omega^-}^2}{M'^2}}\frac{f_{J/\psi} \lambda_{\Omega^-} \lambda_{\widetilde{P}_{csss}}m_{J/\psi}}{(m_{J/\psi}^2 + Q^2)},
\end{eqnarray}
and $\tilde{\Pi}_{a,i}$ are the Borel transformed results of the QCD side corresponding to the results obtained for the first and second currents. These six coupled equations are solved for the unknowns, which are the strong coupling form factors entering the transitions of the negative and positive parity pentaquarks. The strong coupling form factors at $ Q^2=- m_{J/\psi}^2$ are called the strong coupling constants defining the strong decay  under study. As we previously mentioned, the results for $\tilde{\Pi}_{a,i}$ are too long, and therefore we do not provide their explicit expressions here.

\section{Numerical analyses}\label{III}

After extracting the QCD sum rules for the coupling constants in the previous section, we proceed to analyze them numerically. They contain various input parameters such as the masses of the quarks entering the calculations, quark, gluon and mixed quark-gluon condensates, masses of the hadrons, etc. These input parameters are given in Table~\ref{tab:Inputs}.    
\begin{table}[h!]
\begin{tabular}{|c|c|}
\hline\hline
Parameters & Values \\ \hline\hline
$m_{c}$                                     & $1.27\pm 0.02~\mathrm{GeV}$ \cite{Zyla:2020zbs}\\
$m_{s}$                                     & $93^{+11}_{-5}~\mathrm{MeV}$ \cite{Zyla:2020zbs}\\
$\langle \bar{q}q \rangle (1\mbox{GeV})$    & $(-0.24\pm 0.01)^3$ $\mathrm{GeV}^3$ \cite{Belyaev:1982sa}  \\
$\langle \bar{s}s \rangle $                 & $0.8\langle \bar{q}q \rangle$ \cite{Belyaev:1982sa} \\
$m_{0}^2 $                                  & $(0.8\pm0.1)$ $\mathrm{GeV}^2$ \cite{Belyaev:1982sa}\\
$\langle \overline{q}g_s\sigma Gq\rangle$   & $m_{0}^2\langle \bar{q}q \rangle$ \\
$\langle \frac{\alpha_s}{\pi} G^2 \rangle $ & $(0.012\pm0.004)$ $~\mathrm{GeV}^4 $\cite{Belyaev:1982cd}\\
$m_{J/\psi}$                                & $(3096.900\pm0.006)~\mathrm{MeV}$ \cite{Zyla:2020zbs}\\
$m_{\Omega^-}$                              & $( 1672.45\pm 0.29 )~\mathrm{MeV}$ \cite{Zyla:2020zbs}\\
$\lambda_{\Omega^-}$                         & $(0.068 \pm 0.019)~\mathrm{GeV}^3$ \cite{Azizi:2016ddw}\\
$f_{J/\psi}$                                & $(481\pm36)~\mathrm{MeV}$ \cite{Veliev:2011kq}\\
\hline\hline
\end{tabular}%
\caption{Necessary input parameters used in the analyses of the coupling constants.}
\label{tab:Inputs}
\end{table} 
Among the input parameters, we also need the masses and current coupling constants for the considered pentaquark states. These are calculated in Ref.~\cite{Wang:2015wsa} for both of the currents used in this article. Their values are given for negative and positive parities as $m_{P_{csss}}=4.68\pm0.13$~GeV, $\lambda_{P_{csss}}=(6.47\pm 1.10)\times 10^{-3}~\mathrm{GeV}^6$, $m_{\widetilde{P}_{csss}}=4.89\pm0.13$~GeV, $\lambda_{\widetilde{P}_{csss}}=(3.44\pm 0.61)\times 10^{-3}~\mathrm{GeV}^6$ for the first current and $m_{P_{csss}}=4.71\pm0.11$~GeV, $\lambda_{P_{csss}}=(6.84\pm 1.00)\times 10^{-3}~\mathrm{GeV}^6$, $m_{\widetilde{P}_{csss}}=5.40\pm0.08$~GeV, $\lambda_{\widetilde{P}_{csss}}=(12.17\pm 1.28)\times 10^{-3}~\mathrm{GeV}^6$ for the second current~\cite{Wang:2015wsa}. Note that the central values of the masses for the negative parity pentaquark states for both currents are lower than the total mass values of the final states, $m_{J/\psi}+m_{\Omega^-}$. Therefore, in the calculations, considering the errors of these masses, we take their upper values, which are higher than the $m_{J/\psi}+m_{\Omega^-}$.

\begin{figure}[h!]
\begin{center}
\includegraphics[totalheight=5cm,width=7cm]{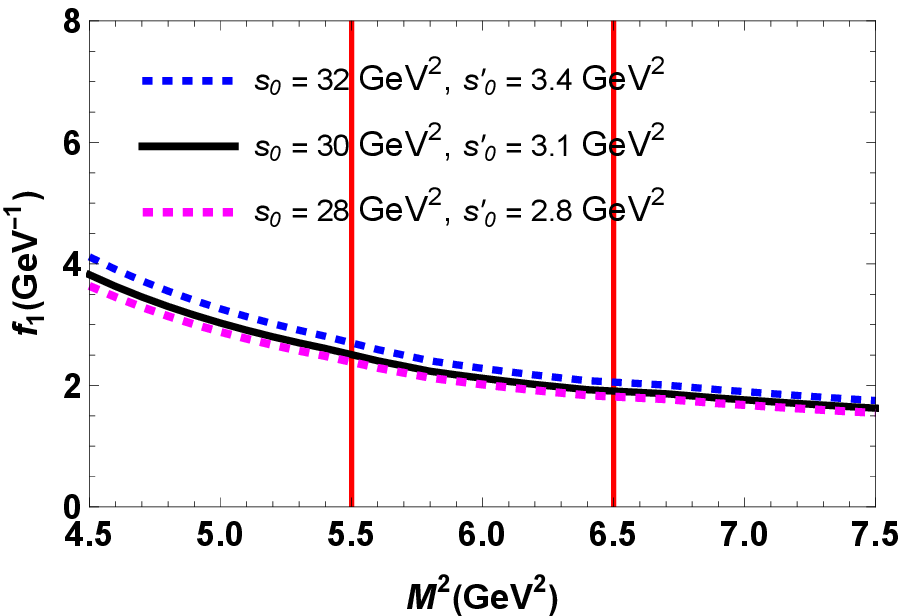}
\includegraphics[totalheight=5cm,width=7cm]{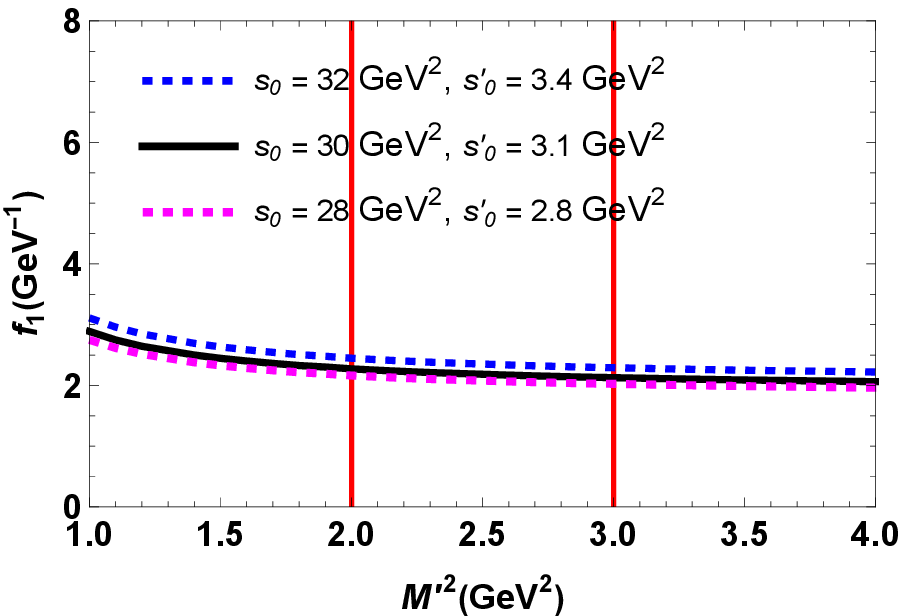}
\end{center}
\caption{\textbf{Left:} Variation of the the strong coupling  form factor $f_1$ obtained from the  first  current as function of $M^2$  at different fixed values of the continuum thresholds,  central value of $M'^2$ and at $Q^2=7.5~\mathrm{GeV^2}$  (the vertical lines are the boundaries of working region for $M^2$.)
\textbf{Right:} The same as the left panel but as a function of  $M'^2$ at central value of $M^2$. }
\label{gr:MsqMpsq}
\end{figure}

In addition, there enter four more auxiliary parameters in the analyses. These are the Borel parameters $M^2$ and $M'^2$, and the threshold parameters, $s_0$ and $s'_0$. The continuum thresholds are selected  such that the analyses include both the resonances under study. To extract the working regions for these parameters, we follow some standard criteria required by the QCD sum rules. For the upper limits of the Borel parameters, the criterion is that contributions coming from the interested states (the two first resonances)   dominate over the higher states and continuum, and for their lower limits the convergence of the OPE (the perturbative contribution exceeds the total nonperturbative contribution and the higher the dimension of the  nonperturbative operator the lower its contribution) is taken into account.  For the pole or two resonances dominance, we require that,
\begin{eqnarray}
PC=\frac{\Pi_{a,i}(s_0,s'_0,M^2,M'^2)}{\Pi_{a,i}(\infty,\infty,M^2,M'^2)},
\end{eqnarray} 
to be  equal or greater than $ 0.5 $. This means that the two resonances under consideration constitute at least $50\%$ of the total contribution. By this way we  fix the upper limits of the  Borel parameters.  As for the lower limits of the Borel parameters, we demand the OPE convergence.  We require that the following parameter that represents the contributions of the higher two dimensions: 
\begin{eqnarray}
R=\frac{\Pi_{a,i}^{(D5+D6)}(s_0,s'_0,M^2,M'^2)}{\Pi_{a,i}(s_0,s'_0,M^2,M'^2)},
\end{eqnarray} 
be equal or smaller than $ 0.05 $, i.e.  the nonperturbative contributions of dimensions five and six constitute maximally the  $5\%$, of the total contributions for each  current and selected structure. Furthermore, in the determined regions, the results are required to be as independent as possible of these variables. Considering all these criteria, the working regions  for the Borel parameters are determined as:
\begin{eqnarray}
5.5~\mbox{GeV}^2\leq M^2\leq 6.5~\mbox{GeV}^2,~~~~~~~~~~2.0~\mbox{GeV}^2\leq M'^2\leq 3.0~\mbox{GeV}^2.
\end{eqnarray}
As for the threshold parameters, they are related to the energies of the excited states at the initial and final channels. Considering again the dominance of the two considered resonances over the higher states and continuum and  the OPE convergence as well as the relatively weak dependence of the results on continuum thresholds, the following working windows are obtained: 
\begin{eqnarray}
28~\mbox{GeV}^2\leq s_0 \leq 32~\mbox{GeV}^2,~~~~~~~~~~2.8~\mbox{GeV}^2\leq s'_0\leq 3.4~\mbox{GeV}^2.
\end{eqnarray}
To see  the dependence of the results on the Borel parameters $M^2$ and $M'^2$ as well as  the thresholds $s_0$ and $s'_0$, we plot,  as an example,  the strong coupling form factor  $f_1$ with respect to the Borel parameters in wide ranges containing their working intervals (bordered by vertical red lines ) at different fixed values of the continuum thresholds and at $Q^2=7.5~\mathrm{GeV^2}$  for the first current of the pentaquark state in figure~\ref{gr:MsqMpsq}. From this figure, we see that the dependence of $f_1$  on the auxiliary parameters in their working intervals is weak. The residual dependencies on $M^2$ and $M'^2$ as well as  the thresholds $s_0$ and $s'_0$ appear as the uncertainties in the presented values. 

Using the working windows of the auxiliary parameters and the values of other  inputs, the behavior of the strong form factors with respect to $Q^2$ can be discussed. 		The strong form factors, obtained in the previous section, give reliable results at positive values of  $Q^2$, however, as previously mentioned, we need their values at $ Q^2=- m_{J/\psi}^2$. The strong form factors obtained at this point are called the strong coupling constants defined the strong decay under study. Hence, we need to extrapolate the results to the negative values of  $Q^2$ using some fit functions. Different fit functions are candidates for this purpose, however, we use the one that best fits to the sum rules predictions at positive values of $Q^2$. The uncertainties related to the fitting procedure is reflected in the presented values, as well. The following fit function best describes the   $Q^2$ behavior of our six strong form factors:
\begin{eqnarray}
g_i(f_i)(Q^2)&=& g_{0}(f_{0})e^{c_1\frac{Q^2}{m_{P_{csss}}^2}+c_2(\frac{Q^2}{m_{P_{csss}}^2})^2},
\end{eqnarray}
where the fitting parameters are given in tables~\ref{tab:FitParam1} and  \ref{tab:FitParam2} for the first and second pentaquark currents, respectively. The values of the strong coupling constants are also presented  at $Q^2=-m_{J/\psi}^2$  in the last column of these tables for both the currents.
\begin{table}[tbp]
\begin{tabular}{|c|c|c|c|c|c|}
\hline\hline
Decay Channel& Coupling Constant  &$ g_0(f_0)$ & $c_1$ & $c_2$ & $g_i(f_i)(-m_{J/\psi}^2) $ \\ \hline\hline
\multirow{3}{*}{ $P_{csss} \rightarrow \Omega^- J/\psi$}  &$g_1~(\mathrm{GeV}^{-1})$&$1.25  $ &$2.46$ & $-0.51$ & $0.41 \pm 0.16$\\ \cline{2-6} 
& $g_2~(\mathrm{GeV}^{-2})$& $-0.23$ & $4.17$ &$3.05$ & $-0.07 \pm 0.03$\\
\cline{2-6} 
& $g_3~(\mathrm{GeV}^{-2})$&$0.83$ & $3.49$ & $-0.04$ &$0.19 \pm 0.14$\\
\hline\hline 
\multirow{3}{*}{ $\widetilde{P}_{csss} \rightarrow \Omega^- J/\psi$}  &$f_1~(\mathrm{GeV}^{-1})$&$0.62  $ &$4.98$ & $-1.21$ & $0.07 \pm 0.01$\\
\cline{2-6}& $f_2~(\mathrm{GeV}^{-2})$&$-0.59$ & $4.58$ & $2.86$ & $-0.15 \pm 0.06$\\
\cline{2-6} 
& $f_3~(\mathrm{GeV}^{-2})$&$1.48$ & $1.37$ & $4.45$ & $1.75 \pm 0.25$\\
\hline\hline 
\end{tabular}%
\caption{ Parameters for the fit functions of
strong coupling constants, $g_1$, $g_2$, and $g_3$ for $P_{csss} \rightarrow \Omega^- J/\psi$ decay; and $f_1$, $f_2$, and $f_3$ for $\widetilde{P}_{csss} \rightarrow \Omega^- J/\psi$  decay  obtained from first current used for the pentaquark state.  The numerical values of the strong coupling constants obtained from the fit functions at $Q^2=-m_{J/\psi}^2$ are also given.} \label{tab:FitParam1}
\end{table}
\begin{table}[tbp]
\begin{tabular}{|c|c|c|c|c|c|}
\hline\hline
Decay Channel& Coupling Constant  &$ g_0(f_0)$ & $c_1$ & $c_2$ & $g_i(f_i)(-m_{J/\psi}^2) $ \\ \hline\hline
\multirow{3}{*}{ $P_{csss} \rightarrow \Omega^- J/\psi$}  &$g_1~(\mathrm{GeV}^{-1})$&$3.33  $ &$-3.11$ & $-25.79$ & $0.15 \pm 0.08$\\ \cline{2-6} 
& $g_2~(\mathrm{GeV}^{-2})$&$26.39$ & $3.73$ & $-1.46$ & $4.40 \pm 0.59$\\
\cline{2-6} 
& $g_3~(\mathrm{GeV}^{-2})$&$-15.19$ & $3.11$ & $0.46$ & $-4.91 \pm 1.21$\\
\hline\hline 
\multirow{3}{*}{ $\widetilde{P}_{csss} \rightarrow \Omega^- J/\psi$}  &$f_1~(\mathrm{GeV}^{-1})$&$6.67  $ &$2.30$ & $-29.71$ & $0.13 \pm 0.07$\\
\cline{2-6}& $f_2~(\mathrm{GeV}^{-2})$&$40.75$ & $5.66$ & $-8.89$ & $1.42 \pm 0.04$\\
\cline{2-6} 
& $f_3~(\mathrm{GeV}^{-2})$&$-26.03$ & $5.52$ & $-8.15$ & $-1.75 \pm 0.39$\\
\hline\hline 
\end{tabular}%
\caption{ Parameters for the fit functions of
strong coupling constants, $g_1$, $g_2$, and $g_3$ for $P_{csss} \rightarrow \Omega^- J/\psi$ decay; and $f_1$, $f_2$, and $f_3$ for $\widetilde{P}_{csss} \rightarrow \Omega^- J/\psi$  decay  obtained from the second current used for pentaquark state.  The values of strong coupling constants obtained from the fit functions at $Q^2=-m_{J/\psi}^2$ are also given.} \label{tab:FitParam2}
\end{table}
The errors  presented for the values of the strong coupling constants  in tables~\ref{tab:FitParam1} and  \ref{tab:FitParam2}  are due to the uncertainties inherited by  the determination of the working intervals of the auxiliary parameters, the errors of the input parameters given in Table~\ref{tab:Inputs} as well as the fitting procedures. For some of the input parameters given in the Table~\ref{tab:Inputs}, such as the mass and condensate of the  strange quark and gluon condensate, there  are also some recent estimates in the literature~\cite{Harnett:2021zug,Dominguez:2014pga}. In our predictions, we consider the effects of the usage of such input values as well  and reflect the outcome in our error estimations.

The final task in this section is to calculate the width of the strong decay channels under consideration. We  find the following decay width formulas:
\begin{eqnarray}
\Gamma &=& \frac{f(m_{P_{csss}},m_{J/\psi},m_{\Omega^-}) }{16\pi m_{P_{csss}}^2}\Bigg[\frac{( (m_{\Omega^-} + m_{P_{csss}})^2-m_{J/\psi}^2 )}{12m_{\Omega^-}^2}\Big(4 g_3^2 m_{J/\psi}^2 [m_{J/\psi}^4 + (m_{\Omega^-}^2 - m_{P_{csss}}^2)^2 + 
    2 m_{J/\psi}^2 (5 m_{\Omega^-}^2 \nonumber\\
    &-& m_{P_{csss}}^2)] 
    + 
 4 g_2 g_3 m_{J/\psi}^2 [m_{J/\psi}^4 - 11 m_{\Omega^-}^4 + 10 m_{\Omega^-}^2 m_{P_{csss}}^2 +
     m_{P_{csss}}^4 - 2 m_{J/\psi}^2 (m_{\Omega^-}^2 + m_{P_{csss}}^2)] + 
 8 g_1^2 [m_{J/\psi}^4 \nonumber\\
 &+&
  (m_{\Omega^-} - m_{P_{csss}})^2 (7 m_{\Omega^-}^2 - 
       4 m_{\Omega^-} m_{P_{csss}} + m_{P_{csss}}^2) - 
    2 m_{J/\psi}^2 (m_{\Omega^-}^2 - 3 m_{\Omega^-} m_{P_{csss}} + m_{P_{csss}}^2)] + 
 g_2^2 [m_{J/\psi}^6 \nonumber\\
 &-& 2 m_{J/\psi}^4 (3 m_{\Omega^-}^2 + m_{P_{csss}}^2) + 
    m_{J/\psi}^2 (3 m_{\Omega^-}^2 + m_{P_{csss}}^2)^2 + 
    8 (m_{\Omega^-}^3 - m_{\Omega^-} m_{P_{csss}}^2)^2] - 
 8 g_1 m_{\Omega^-} \big[12 g_3 m_{J/\psi}^2 m_{\Omega^-} (m_{\Omega^-} \nonumber\\
 &-& m_{P_{csss}}) + 
    g_2 [m_{J/\psi}^4 - (m_{\Omega^-} - m_{P_{csss}})^2 (5 m_{\Omega^-}^2 + 
          4 m_{\Omega^-} m_{P_{csss}} - m_{P_{csss}}^2) - 
       2 m_{J/\psi}^2 (m_{\Omega^-}^2 + m_{P_{csss}}^2)]\big]\Big)\Bigg],
\label{Eq:DWnp}
\end{eqnarray}
for the negative parity pentaquark and
\begin{eqnarray}
\Gamma &=& \frac{f(m_{\widetilde{P}_{csss}},m_{J/\psi},m_{\Omega^-}) }{16\pi m_{\widetilde{P}_{csss}}^2}\Bigg[\frac{( (m_{\Omega^-} - m_{\widetilde{P}_{csss}})^2-m_{J/\psi}^2 )}{12m_{\Omega^-}^2}\Big( (4 f_3^2 m_{J/\psi}^2 [m_{J/\psi}^4 + (m_{\Omega^-}^2 - 
         m_{\widetilde{P}_{csss}}^2)^2 + 2 m_{J/\psi}^2 (5 m_{\Omega^-}^2 \nonumber\\
         &-&
          m_{\widetilde{P}_{csss}}^2)] + 
    4 f_2 f_3 m_{J/\psi}^2 [m_{J/\psi}^4 - 11 m_{\Omega^-}^4 + 
       10 m_{\Omega^-}^2 m_{\widetilde{P}_{csss}}^2 + m_{\widetilde{P}_{csss}}^4 - 
       2 m_{J/\psi}^2 (m_{\Omega^-}^2 + m_{\widetilde{P}_{csss}}^2)] + 
    8 f_1^2 [m_{J/\psi}^4\nonumber\\
    & -& 
       2 m_{J/\psi}^2 (m_{\Omega^-}^2 + 3 m_{\Omega^-} m_{\widetilde{P}_{csss}} + 
          m_{\widetilde{P}_{csss}}^2) + (m_{\Omega^-} + m_{\widetilde{P}_{csss}})^2 (7 m_{\Omega^-}^2 + 
          4 m_{\Omega^-} m_{\widetilde{P}_{csss}} + m_{\widetilde{P}_{csss}}^2)] + 
    f_2^2 [m_{J/\psi}^6\nonumber\\
    & -&
     2 m_{J/\psi}^4 (3 m_{\Omega^-}^2 + m_{\widetilde{P}_{csss}}^2) + 
       m_{J/\psi}^2 (3 m_{\Omega^-}^2 + m_{\widetilde{P}_{csss}}^2)^2 + 
       8 (m_{\Omega^-}^3 - m_{\Omega^-} m_{\widetilde{P}_{csss}}^2)^2] - 
    8 f_1 m_{\Omega^-} \big[12 f_3 m_{J/\psi}^2 m_{\Omega^-} (m_{\Omega^-} \nonumber\\
    &+&
     m_{\widetilde{P}_{csss}}) + 
       f_2 [m_{J/\psi}^4 - (m_{\Omega^-} + m_{\widetilde{P}_{csss}})^2 (5 m_{\Omega^-}^2 - 
             4 m_{\Omega^-} m_{\widetilde{P}_{csss}} - m_{\widetilde{P}_{csss}}^2) - 
          2 m_{J/\psi}^2 (m_{\Omega^-}^2 + m_{\widetilde{P}_{csss}}^2))]\big]\Big)\Bigg],
\label{Eq:DWpp}
\end{eqnarray}
for the positive parity pentaquark states with 
\begin{eqnarray}
f(x,y,z)&=&\frac{1}{2x}\sqrt{x^4+y^4+z^4-2x^2y^2-2x^2z^2-2y^2z^2}.\label{functionf}
\end{eqnarray} 
We obtain the numerical values of the corresponding widths of the considered decay channels as 
\begin{eqnarray}
\Gamma(P_{csss} \rightarrow J/\psi \Omega^-)&=& 201.4\pm 82.5~\mathrm{MeV},
 \label{Eq:DWnp1}
\end{eqnarray}
for the negative parity pentaquark, and
\begin{eqnarray}
\Gamma(\widetilde{P}_{csss} \rightarrow J/\psi \Omega^-)&=& 316.4\pm 107.8~\mathrm{MeV},
 \label{Eq:DWpp1}
\end{eqnarray}
for the positive parity pentaquark states when the first interpolating current is used. And,
\begin{eqnarray}
\Gamma(P_{csss} \rightarrow J/\psi \Omega^-)&=& 252.5\pm 116.7~\mathrm{MeV},
 \label{Eq:DWnp1}
\end{eqnarray}
for the negative parity pentaquark, and
\begin{eqnarray}
\Gamma(\widetilde{P}_{csss} \rightarrow J/\psi \Omega^-)&=& 361.1\pm 98.4~\mathrm{MeV},
 \label{Eq:DWpp1}
\end{eqnarray}
for the positive parity pentaquark states when the second interpolating current is used.

\section{Summary and conclusion}\label{Sum} 

The improvements reached in time in both the experimental and analyses techniques  have led us to observations of new particles including the exotic states. The indications for possible similar future observations motivates the theoretical studies supplying information for such experimental investigations. In recent years, a pentaquark state with a single strange quark, $P_{cs}(4459)^0$, was reported by the LHCb Collaboration~\cite{LHCb:2020jpq}. This observation attained in the $J/\psi \Lambda$ invariant mass distribution of the $\Xi_b^-\rightarrow J/\psi \Lambda K^-$ also indicates the possible existence of the pentaquark states with double or triple strange quark content, which may be observed in the near future. This expectation has motivated us to study the possible double or triple pentaquark states. We considered a possible double strange pentaquark state in Ref.~\cite{Azizi:2021pbh}. In the present study, we investigated the strong transitions, $P_{csss}(\tilde{P}_{csss}) \rightarrow J/\psi \Omega^-$, for a possible triple strange pentaquark state with negative (positive) parity within the framework of the QCD sum rules considering two different interpolating currents with spin-parity $J^P=\frac{1}{2}^-$. The mass predictions for the considered state were obtained using the same currents in Ref.~\cite{Wang:2015wsa}, and we used these results in our decay analyses. 

The strong decay analyses may supply effective support besides the mass predictions for fixing the possible quantum numbers of these particles in their future observations. In some analyses of currently observed pentaquark states, their decays were also considered to fix the properties of these particles besides the spectral analyses, since spectral analyses might not suffice to do that. With this drive, in this work, we calculated the corresponding strong  coupling constants for the decays, $P_{csss}(\tilde{P}_{csss}) \rightarrow J/\psi \Omega^-$, and using them we obtained the corresponding decay widths considering the two mentioned interpolating currents. Though the currents have negative parity, they can create states both with the negative and positive parities. And therefore, we obtained the strong coupling constants  $g_{i}$ and $f_{i}$  for negative and positive parity states, respectively, at $Q^2=-m_{J/\psi}^2$ and using them we calculated their corresponding decay widths for each current. These decay widths are obtained for the negative and positive parity states as $\Gamma(P_{csss} \rightarrow J/\psi \Omega^-)=201.4\pm 82.5~\mathrm{MeV}$ and  $\Gamma(\widetilde{P}_{csss} \rightarrow J/\psi \Omega^-)=316.4\pm 107.8~\mathrm{MeV}$, respectively, using the first current, as well  as  $\Gamma(P_{csss} \rightarrow J/\psi \Omega^-)=252.5\pm 116.7~\mathrm{MeV}$ and  $\Gamma(\widetilde{P}_{csss} \rightarrow J/\psi \Omega^-)=361.1\pm 98.4~\mathrm{MeV}$, respectively, using the second current.

In the near future, for the possible observation of such triple strange pentaquark states, the results of this work may provide valuable information through the comparison of the obtained results with the experimental observations. These comparisons may help in decisive fixing of  the physical properties of these states.

\section*{ACKNOWLEDGEMENTS}
K. Azizi is thankful to Iran Science Elites Federation (Saramadan)
for the partial  financial support provided under the grant number ISEF/M/401385.



\end{document}